\RequirePackage{snapshot}% <-- Optional, for use with bundledoc only 
\documentclass[aps,pra,twocolumn,superscriptaddress,nofootinbib,showpacs]{revtex4-1}

\usepackage[english]{babel}
\usepackage{amsmath}
\usepackage{amsfonts}
\usepackage{amssymb}
\usepackage{amsthm}
\usepackage{graphicx}
\usepackage{braket}
\usepackage{hyperref}
\usepackage{todonotes}
\usepackage{soul}

\newcommand{\ham}{\mathcal{H}}
\newcommand{\sigmat}{\boldsymbol{\sigma}}
\newcommand{\Piop}{\mathbf{\Pi}}
\newcommand{\aop}{\hat{a}}
\newcommand{\aopd}{\hat{a}^\dagger}
\newcommand{\In}{\ensuremath{(\mathrm{in})}}
\newcommand{\Out}{\ensuremath{(\mathrm{out})}}
\newcommand{\Rabi}{\ensuremath{\Omega_\mathrm{R}}}

\begin{document}
	
\title{Driven quantum tunneling and pair creation with graphene Landau levels}
\author{Denis Gagnon}
\email{denis.gagnon@uwaterloo.ca}
\author{Fran\c{c}ois Fillion-Gourdeau}
\affiliation{Universit\'e du Qu\'ebec, INRS--\'Energie, Mat\'eriaux et T\'el\'ecommunications, Varennes, Qu\'ebec, Canada, J3X 1S2}
\affiliation{Institute for Quantum Computing, University of Waterloo, Waterloo, Ontario, Canada, N2L 3G1}
\author{Joey Dumont}
\affiliation{Universit\'e du Qu\'ebec, INRS--\'Energie, Mat\'eriaux et T\'el\'ecommunications, Varennes, Qu\'ebec, Canada, J3X 1S2}
\author{Catherine Lefebvre}
\author{Steve MacLean}
\email{steve.maclean@uwaterloo.ca}
\affiliation{Universit\'e du Qu\'ebec, INRS--\'Energie, Mat\'eriaux et T\'el\'ecommunications, Varennes, Qu\'ebec, Canada, J3X 1S2}
\affiliation{Institute for Quantum Computing, University of Waterloo, Waterloo, Ontario, Canada, N2L 3G1}
\date{\today}

\begin{abstract}
	 
Driven tunneling between graphene Landau levels is theoretically linked to the process of pair creation from vacuum, a prediction of quantum electrodynamics (QED).
Landau levels are created by the presence of a strong, constant, quantizing magnetic field perpendicular to a graphene mono-layer.
Following the formal analogy between QED and the description of low-energy excitations in graphene, solutions of the fully interacting Dirac equation are used to compute electron-hole pair creation driven by a circularly or linearly polarized field.
This is achieved via the coupled channel method, a numerical scheme for the solution of the time-dependent Dirac equation in the presence of bound states. 
The case of a monochromatic driving field is first considered, followed by the more realistic case of a pulsed excitation.
We show that the pulse duration yields an experimental control parameter over the maximal pair yield.
Orders of magnitude of the pair yield are given for experimentally achievable magnetic fields and laser intensities weak enough to preserve the Landau level structure.

\end{abstract}


\maketitle

\section{Introduction}

Mono-layer graphene, a planar crystal of carbon atoms arranged on a honeycomb lattice, has been at the heart of condensed matter research for the last decade \cite{CastroNeto2009}.
It has been proposed to use graphene in electronic devices, for example in ballistic transistors \cite{Geim2007}, topological insulators \cite{Kindermann2015} and nano-pore sensors \cite{Puster2013}.
The potential contribution of graphene to integrated optical and optoelectronic devices, such as photo-detectors \cite{Schall2014}, phase modulators \cite{Miao2015} and saturable absorbers for micro-lasers\cite{Canbaz2015}, has also been demonstrated. 
Besides this wealth of practical applications, the low Fermi velocity of graphene ($v_F = 10^6$ m/s) and its linear energy dispersion has led researchers to propose its use as a quantum electrodynamics (QED) test bench \cite{Semenoff1984, Katsnelson2007, Gusynin2007, Geim2007}.
The formal analogy between QED and graphene stems from the fact that the dynamics of low-energy excitations near the $K^\pm$ corners of the Brillouin zone of graphene (also called Dirac points) are governed by the Dirac equation, much like electrons in relativistic quantum mechanics.
Based on this analogy, we will use the terminology ``graphene QED'' in this article to refer to relativistic quantum mechanics in graphene \cite{Fillion-Gourdeau2015}.
A key difference between usual QED and graphene QED is that the Dirac Hamiltonian governing graphene quasiparticle dynamics does not involve a mass term.
The realization that electrons and holes in graphitic materials may behave like massless Dirac fermions has motivated theoreticians and experimentalists to extend celebrated predictions of QED to graphene.
These predictions include Klein tunneling \cite{Stander2009} and electron-positron pair production from vacuum \cite{Allor2008, Lewkowicz2011}, the latter of which is the focus of the present article.

The quest for the observation of pair production from vacuum either through the Schwinger effect or multiphoton processes has driven unprecedented developments in the high-power lasers community \cite{DiPiazza2012}.
Nevertheless, the critical field strength required to observe pair production in usual QED ($E_0 = 1.3 \times 10^{18}$ V/m) is still orders of magnitude greater than what state-of-the art lasers can achieve \cite{Narozhny2015}.
This further motivates the use of graphene as a QED analogue, since the energy scales required to observe the equivalent mechanisms are much lower in graphene than in usual QED.
In fact, because Dirac fermions in graphene are massless, there is no exponential suppression of the Schwinger mechanism, implying that non-perturbative pair production occurs for any applied field strength \cite{Fillion-Gourdeau2015}.
Pair production in graphene driven by a constant \cite{Lewkowicz2009, Dora2010, Rosenstein2010} or time-dependent \cite{Klimchitskaya2013} spatially homogeneous electric field has already been studied theoretically and linked to its transport properties \cite{Gavrilov2012}.
The density of produced pairs can be related to solutions of the time-dependent Dirac equation in the presence of an applied field via the Schwinger-Keldysh formalism \cite{Gelis2006, Fillion-Gourdeau2015, Gelis2015}.

In this article, we extend this analogy between Dirac fermion dynamics and pair creation to the case of graphene in a strong perpendicular magnetic field.
Since the presence of the quantizing magnetic field condenses the electron orbits into bound states called \emph{Landau levels} \cite{Goerbig2011}, fundamental differences exist between the magnetized and non-magnetized case.
In particular, the presence of the $B$-field breaks translational symmetry, and momentum is no longer a good quantum number, requiring a slightly modified second quantization procedure to compute the number of pairs.
Graphene Landau levels (LLs) also exhibit a large macroscopic degeneracy, which is proportional to the magnitude of the quantizing $B$-field \cite{Gusynin2007, Goerbig2011}.
We show theoretically and numerically that the pair yield can be increased by applying a stronger magnetic field, because the number of fermions that can be ``packed'' in a given graphene LL increases accordingly.

This paper is organized as follows.
In Section \ref{sec:driven}, we theoretically describe the driven tunneling between LLs, starting with a review of the canonical quantization of graphene in a magnetic field (Section \ref{sec:canonical}).
We then shift to the interaction picture and describe the selection rules specific to graphene in Section \ref{sec:interaction}.
The central contribution of this article is the formal link between solutions of the time-dependent Dirac equation and pair production in Landau quantized graphene.
This link is made in Section \ref{sec:pair} using second quantized theory.
Numerical solutions of the time-dependent Dirac equation via the coupled channel method are subsequently presented in Section \ref{sec:numerical} for two types of circularly polarized driving fields: a monochromatic excitation and a pulse containing a finite number of carrier cycles.
These results are then interpreted in terms of an electron-hole pair density.
It is shown that the applied pulse duration provides a control parameter that allows one to maximize the number of produced pairs.
The case of linear polarization, which is associated to an increase of the pair yield, is discussed in Section \ref{sec:linear}, and the conclusion is found in Section \ref{sec:conclusion}.

\section{Driven tunneling between graphene Landau levels}
\label{sec:driven}
Electronic properties of graphene in a quantizing magnetic field have been the object of several theoretical review articles \cite{CastroNeto2009,Goerbig2011}.
On the experimental side, placing graphene in a strong magnetic field has led researchers to several major milestones.
Most notably, the observation of the integer quantum Hall effect by Novoselov \emph{et al.} was the first demonstration of the massless nature of Dirac fermions in graphene \cite{Novoselov2005}.
Magnetized graphene is also expected to be of great use for applications related to quantum information \cite{Tokman2013}. 
In this work, we aim to establish the potential of magnetized graphene as a QED analogue by relating Landau level dynamics to time-dependent pair creation (Section \ref{sec:pair}).
For this purpose we first proceed to review Landau quantization (Section \ref{sec:canonical}) and quantum optics of LLs in the presence of a time-dependent driving electric field (Section \ref{sec:interaction}).

\subsection{Landau quantization}
\label{sec:canonical}
\begin{figure}
	\centering
	\includegraphics[]{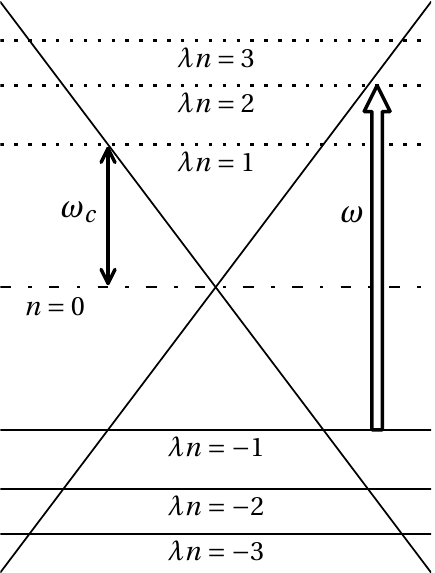}
	\caption{Landau levels (LLs) in graphene in the presence of a quantizing magnetic field. LL energies are $\epsilon_{\lambda n} \propto \lambda \sqrt{Bn}$, where $n$ is non-negative integer. In the case of neutral graphene, all hole-like states are filled (solid lines), and the zero-energy LL is half-filled (dashed-dotted line).
		The Dirac cone describing the linear relationship between energy and momentum in the case of free fermions is shown for convenience.
		Throughout this paper, it will be assumed that an incident laser excitation drives the transition between LLs -1 and 2, as indicated by the white arrow.}\label{fig:levels}
\end{figure}

Consider Dirac fermions in a graphene mono-layer in the presence of a quantizing magnetic field, which is uniform in space and directed perpendicular to the graphene plane.
The dynamics of the fermions are governed by the following $(2 \times 2)$ low energy Hamiltonian (we use units such that $\hbar = 1$) \cite{Goerbig2011}
\begin{equation}\label{eq:hamiltonianb}
\ham_{\xi}^\mathbf{B} = \xi v_F \sigmat \cdot \Piop^\mathbf{B},
\end{equation}
where $\xi = \pm 1$ is the valley pseudospin index, $v_F$ is the Fermi velocity in graphene, $\sigmat = (\sigma_x, \sigma_y)$ is a vector of Pauli matrices representing the sublattice pseudospin, $\Piop^\mathbf{B} \equiv \mathbf{p} + e \mathbf{A}^\mathbf{B}(\mathbf{r})$ is the canonical momentum, $\mathbf{p}$ is the fermion momentum around the $K^\xi$ points and $-e < 0$ is the electron charge.
The magnetic field is related to the vector potential via the usual relation $\mathbf{B} = \nabla \times \mathbf{A}^\mathbf{B}(\mathbf{r})$.
The gauge choice is arbitrary\footnote{For instance, one could use the Landau gauge: $\mathbf{A}^\mathbf{B} = (-By,0)$}, as long as the canonical momentum satisfies the following commutation relation \cite{Goerbig2011}
\begin{equation}\label{eq:canonical}
[ \Pi_x^\mathbf{B}, \Pi_y^\mathbf{B}] = - \frac{i}{l_B^2},
\end{equation}
where the magnetic length is defined as $l_B \equiv \sqrt{1 /eB}$, with $B$ the magnitude of the quantizing magnetic field.
Assuming that intervalley coupling can be neglected, the following implicit two-spinor representation can be employed
\begin{equation}
\ket{\varphi_{\xi = +}} = \begin{pmatrix} \varphi_{A,+} \\ \varphi_{B,+} \end{pmatrix} \qquad, \qquad 
\ket{\varphi_{\xi = -}} = \begin{pmatrix} \varphi_{B,-} \\ \varphi_{A,-} \end{pmatrix},
\end{equation}
where the first subscript stands for the sublattice pseudospin.
The system is formally equivalent to a quantum harmonic oscillator because of the canonical commutation relation \eqref{eq:canonical}.
Consequently, we can use the same calculation technique.
The eigenvalues and eigenfunctions of \eqref{eq:hamiltonianb} may be found by introducing the usual ladder operators and their eigenstates \cite{Goerbig2011}
\begin{equation}
\aop = \frac{l_B}{\sqrt{2}} \big( \Pi^\mathbf{B}_x - i \Pi^\mathbf{B}_y \big), \qquad \aopd = \frac{l_B}{\sqrt{2}} \big( \Pi^\mathbf{B}_x + i \Pi^\mathbf{B}_y \big),
\end{equation}
\begin{equation}
\aop \ket{n} = \sqrt{n} \ket{n-1} , \qquad \aopd \ket{n} = \sqrt{n+1} \ket{n+1}.
\end{equation}
The ladder operator eigenstates satisfy $\braket{n| n'} = \delta_{nn'}.$
The Hamiltonian can be rewritten as
\begin{equation}\label{eq:hamiltonianb2}
\ham_{\xi}^\mathbf{B} = \xi \omega_c 
\begin{pmatrix} 0 & \aop \\ \aopd & 0 \end{pmatrix},
\end{equation}
where we introduce the \emph{cyclotron frequency}
\begin{equation}
\omega_c \equiv \sqrt{2} \frac{v_F}{l_B}.
\end{equation}
The eigenvalue spectrum of \eqref{eq:hamiltonianb} consists of discrete states with energies given by
\begin{equation}
\epsilon_{\lambda n} = \lambda \sqrt{n} \omega_c.
\end{equation}
These discrete energy levels are called \emph{Landau levels} (LLs).
The solutions are now labeled using the band index $\lambda = \pm$ (positive for the conduction band, negative for the valence band) and the LL index $n \in \mathbb{N}_0$ (see Fig. \ref{fig:levels}).
Up to a normalization factor, one can write the following spinor solution for the zeroth LL ($n=0$) \cite{Goerbig2011}
\begin{equation}\label{eq:zll}
\ket{\varphi_{0; \xi} } = 
\begin{pmatrix}
0 \\
\xi \ket{0}
\end{pmatrix}.
\end{equation}
The two-spinor for all other levels ($n \neq 0$) reads
\begin{equation}\label{eq:twospinor}
\ket{\varphi_{\lambda n; \xi}} = \frac{1}{\sqrt{2}}
\begin{pmatrix}
\ket{n-1} \\
\lambda \xi \ket{n}
\end{pmatrix}.
\end{equation}
 
\subsection{Quantum optics of Landau levels}
\label{sec:interaction}
Suppose that a homogeneous but time-dependent electric field is applied parallel to the magnetized graphene layer between $t=0$ and $t=T$.
Quasi-particles in the layer are described by the following low energy Hamiltonian
\begin{equation}\label{eq:hamiltonianbe}
\ham_{\xi} = \xi v_F \sigmat \cdot \Piop(\mathbf{r},t),
\end{equation}
where
\begin{equation}\label{eq:potentialt}
\Piop(\mathbf{r},t) =
\begin{cases}
\Piop^\mathbf{B}(\mathbf{r}), \qquad & t < 0 \\
\Piop^\mathbf{B}(\mathbf{r}) + e \mathbf{A}^\mathbf{E}(t), \qquad & t \in [0,T] \\
\Piop^\mathbf{B}(\mathbf{r}) + e \mathbf{A}^\mathbf{E}(T), \qquad & t > T.
\end{cases}
\end{equation}
The vector potential is chosen such that $\mathbf{A}^\mathbf{E}(t \leq 0) = 0$ and is related to the applied electric field by
\begin{equation}
\mathbf{E}(t) = - \partial_t \mathbf{A}^\mathbf{E}(t).
\end{equation}
The Hamiltonian can be split into a time-independent magnetic component and a time-dependent electric component:
\begin{equation}
\ham_{\xi}(t) = \ham^{\mathrm{B}}_{\xi} + \ham^{\mathrm{int}}_{\xi}(t),
\end{equation}
where the eigenstates of $\ham^{\mathrm{B}}_{\xi}$ are given by Eqs. (\ref{eq:zll}--\ref{eq:twospinor}) and the interaction part is defined as \cite{Wang2015}
\begin{equation}
\ham^{\mathrm{int}}_{\xi}(t) = \xi e v_F \sigmat \cdot \mathbf{A}^\mathbf{E}(t).
\end{equation}
To obtain a solution to this problem, one can expand $\ket{\psi (t)}$ on the basis $\lbrace \ket{\varphi_{\lambda n; \xi}} \rbrace$.
This is useful since, as shown in Section \ref{sec:canonical}, the eigenstates of $\ham^{\mathrm{B}}_{\xi}$ are known.
The basis expansion reads \cite{Sakurai1993, Cohen-Tannoudji2007}
\begin{equation}\label{eq:ansatz}
%\begin{aligned}
\ket{\psi_\xi (t)} =  \sum_{\substack{\lambda n \\ (n \neq 0)}} b_{\lambda n; \xi}(t) e^{-i \epsilon_{\lambda n} t } \ket{\varphi_{\lambda n; \xi}} + b_{0; \xi}(t) \ket{\varphi_{0; \xi}} .
%\end{aligned}
\end{equation}
The time-evolution of the $b_{\lambda n; \xi}(t)$ coefficients is solely dictated by the interaction Hamiltonian, as can be shown from the Dirac equation:
\begin{equation}\label{eq:dirac2}
i \partial_t \ket{\psi_\xi (t)} = \left[\ham^{\mathrm{B}}_{\xi} + \ham^{\mathrm{int}}_{\xi}(t) \right] \ket{\psi_\xi (t)}.
\end{equation}
Substituting Eq. \eqref{eq:ansatz} in Eq. \eqref{eq:dirac2} and projecting the resulting equation on $\bra{\varphi_{\lambda n; \xi}}$ yields
\begin{equation}\label{eq:ode21}
\begin{aligned}
i \dot{b}_{\lambda n; \xi}(t) = 
& \sum_{\substack{\lambda' n' \\ (n' \neq 0)}} b_{\lambda' n'; \xi}(t)  H^{\mathrm{int},\xi}_{\lambda n, \lambda' n'}(t)
e^{i \omega_{\lambda n, \lambda' n'}t} \\
& + b_{0;\xi}(t) H^{\mathrm{int},\xi}_{\lambda n, 0}(t) e^{i \omega_{\lambda n, 0}t}, \qquad (n \neq 0)  \\
\end{aligned}
\end{equation}
where
\begin{equation}
\omega_{\lambda n, \lambda' n'} \equiv \epsilon_{\lambda n} - \epsilon_{\lambda' n'} = \omega_c \left(\lambda \sqrt{n} - \lambda' \sqrt{n'} \right),
\end{equation}
and where we use the following shorthand notation for matrix elements
\begin{equation}\label{eq:matrix}
H^{\mathrm{int},\xi}_{\lambda n, \lambda' n'}(t) \equiv \braket{\varphi_{\lambda n;\xi} | \ham^{\mathrm{int}}_{\xi}(t) | \varphi_{\lambda' n';\xi}}.
\end{equation}

\begin{figure}
	\centering
	\includegraphics[]{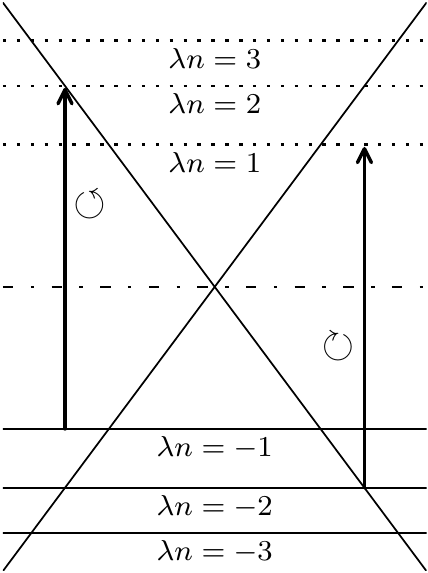}
	\caption{Allowed transitions with frequency $\omega = \omega_c(\sqrt{2} + 1)$ and their respective handedness.
		Transitions from LL indices $n \rightarrow n + 1$ are associated with the absorption of RHP photons, and transitions from $n \rightarrow n - 1$ with the absorption of LHP photons \cite{Abergel2007}.}\label{fig:transitions}
\end{figure}

For the needs of this study let us consider a field of the general form
\begin{equation}\label{eq:potential}
\mathbf{A}^\mathbf{E}(t) = A_0 \left\lbrace G_x(t) \hat{\mathbf{e}}_x + G_y(t)\hat{\mathbf{e}}_y \right\rbrace,
\end{equation}
where the $G_i(t)$ are dimensionless functions of time and $A_0$ is a real number related to the field amplitude.
This describes a time-dependent electric field parallel to the graphene layer.
Substituting Eq. \eqref{eq:potential} in Eq. \eqref{eq:matrix}, using Eqs. \eqref{eq:zll} and \eqref{eq:twospinor} and plugging the resulting matrix elements in Eq. \eqref{eq:ode21}, one obtains
\begin{subequations}\label{eq:ode3}
\begin{equation} 
\begin{aligned}
& i \dot{b}_{\lambda n; \xi}(t) = \\ &\frac{\Rabi}{2} \sum_{\lambda'} \bigg\lbrace   \lambda b_{\lambda' n+1; \xi}(t) G^{+}(t) e^{i \omega_{\lambda n, \lambda' n + 1}t} \\
+ & (1 - \delta_{n-1,0}) \lambda' b_{\lambda' n-1; \xi}(t) G^{-}(t) e^{i \omega_{\lambda n, \lambda' n - 1}t} \bigg\rbrace \\
+ & \delta_{n-1,0} \frac{\Rabi}{\sqrt{2}} b_{0;\xi}(t)  G^{-}(t) e^{i \omega_{\lambda 1, 0}t}, \qquad (n \neq 0).
\end{aligned}
\end{equation}
The differential equation for the zeroth LL (ZLL, $n=0$) is
\begin{equation}
i \dot{b}_{0;\xi} (t) = \frac{\Rabi}{\sqrt{2}} \sum_{\lambda} b_{\lambda 1;\xi} (t) G^{+}(t) e^{i \omega_{0, \lambda 1}t}.
\end{equation}
\end{subequations}
where $G^\pm(t) = G_x(t) \pm i G_y(t)$, and we have defined the Rabi frequency
\begin{equation}
\Rabi \equiv e v_F A_0.
\end{equation}
As revealed in previous studies \cite{Tokman2013, Wang2015}, the form of the system of Eqs. \eqref{eq:ode3} indicates that every level characterized by the quantum number $n$ is coupled to the two adjacent levels with quantum numbers $n \pm 1$, both in the valence ($\lambda = -$) and the conduction ($\lambda = +$)  band, for a total of 4 allowed transitions (except for LLs $n = 1$ which are coupled to 3 levels, and LL $n=0$ which is coupled to 2 levels).
Using the rotating wave approximation (RWA, see Section \ref{sec:2level}) it can be shown that every allowed transition from the valence to the conduction band is either associated with the absorption of a right-handed or left-handed polarized excitation \cite{Abergel2007, Tokman2013, Yao2012, Yao2013, Wang2015}.
As described in Section \ref{sec:pair}, this type of upward transition is directly linked to pair creation.
Figure \ref{fig:transitions} illustrates the fact that transitions from LL indices $n \rightarrow n + 1$ are associated with the absorption of right-handed polarized (RHP) photons, and transitions from $n \rightarrow n - 1$ are associated with the absorption of LHP photons \cite{Abergel2007, Yao2013}.

These peculiar transition rules specific to the 2D relativistic Dirac Hamiltonian (two intraband and two interband transitions) can be contrasted to the non-relativistic case of the 2D electron gas (2DEG), which can be realized using semiconductor heterostructures.
The selection rules specific to graphene make it possible to optically address a single transition by tailoring the laser frequency and polarization.
In contrast, the only dipole allowed transitions in a 2DEG are those from LLs $n$ to $n + 1$, and all have the same transition energy irrespective of the LL index \cite{Goerbig2011}.
This is due to the equidistant level structure of the 2DEG.
Furthermore, Landau quantization in graphene \emph{reduces} the impact of many-body effects, i.e. Auger scattering, whereas in a non-relativistic 2DEG, Auger scattering is \emph{enhanced} \cite{Plochocka2009}.
Acoustic phonon scattering is also weaker in graphene than in an ordinary 2DEG \cite{Wendler2014}.
In short, the peculiar selection rules of graphene and its higher carrier lifetime motivate its choice over a 2DEG for pair creation studies.

For definiteness, we shall only consider right-handed circularly polarized (RHP) excitations (i.e. an electric field vector rotating counter-clockwise around the quantizing magnetic field vector) for the remainder of this article unless where noted.
As seen from the semi-classical picture (Fig. \ref{fig:transitions}), this excitation drives the transition between LLs -1 and 2.
A rotating electric field could be generated either using electrodes or counter-propagating circularly polarized lasers: in the latter configuration, there is a cancellation of the laser induced magnetic field at anti-nodes of the generated standing wave.
Furthermore, since a linearly polarized excitation can be viewed as the superposition of a right-handed and a left-handed polarized (LHP) excitation, linear polarization results can be straightforwardly interpreted in terms of the circular case.
Because every allowed transition frequency $\omega_{\lambda n, \lambda' n'}$ is associated to two graphene LL transitions of different handedness, using a linearly polarized excitation basically amounts to driving two transitions simultaneously instead of only one.
For this reason, we relegate discussion of pair production with linearly polarized excitations to Section \ref{sec:linear}.

\subsubsection{Numerical method}
To obtain the quasi-particle dynamics in the presence of LLs driven by a in-plane electric field, one has to solve the system of coupled ordinary differential equations (ODEs) given by Eqs. \eqref{eq:ode3}.
Except for some special cases (see Section \ref{sec:2level}), there exists no analytic solution to the system.
It is however amenable to a numerical solution via the coupled-channel method (CCM), which is targeted at quantum systems interacting with external fields.
The CCM consists in solving the time-dependent Dirac or Schr\"{o}dinger equation by expanding the solution on a eigenstate basis \cite{Wang2008}.
For the problem at hand, this is realized by the system of Eqs. \eqref{eq:ode3}, where the basis set is composed of the unperturbed LLs.
In practical implementations, only a finite number of energy levels are considered, say $2 \mathcal{N} + 1$ LLs, including $\mathcal{N}$ levels above and $\mathcal{N}$ levels below the ZLL.
The resulting system of ordinary differential equations is then solved numerically. 
The description of the numerical method along with technical details are relegated to Appendix \ref{sec:ccm}.

It should be noted that quasi-particle dynamics in graphene in the presence of a time-dependent driving field can also be described using Floquet theory.
In the non-magnetized case, a circularly polarized excitation can be shown to induce an intensity dependent gap in Dirac cones \cite{Oka2009}.
In the Landau-quantized case, Floquet theory applied to a similar excitation yields a photoinduced LL modulation as described by L\'opez \emph{et al.} \cite{Lopez2015}
Time-dependent results are also obtained in the latter paper in the low Rabi frequency (i.e. low intensity) limit.
In contrast, the CCM used in the present article describes the LL dynamics for \emph{any} value of the Rabi frequency.
This could be achieved using a fully numerical Floquet calculation, although the usefulness of this treatment is appropriate to the high frequency regime \cite{Cayssol2013}, i.e. for $\omega/\Rabi \gg 1$ and $\omega / \omega_c \gg 1$.

\subsubsection{Two-level approximation}
\label{sec:2level}
In this subsection, we describe briefly the use of the rotating wave approximation (RWA) to reduce Eqs. \eqref{eq:ode3} to a two-level system \cite{Boyd2003}.
The motivation of this exercise is two-fold.
First, it allows one to show that Rabi oscillations (a hallmark of driven quantum tunneling \cite{Grifoni1998}) play a key part in graphene LL dynamics.
Indeed, the phenomenon of Rabi oscillations in graphene was investigated by several groups \cite{Dora2009, Lopez2015}.
Second, the results obtained via the two-level approximation may be compared \emph{a posteriori} with the results obtained via the CCM, since a good agreement between the two approaches is expected in the range of validity of the RWA (moderate values of $\Rabi/ \omega_c$).
For a broadband pulse with spectral width $\delta \omega$, an additional restriction on the validity of the two-level approximation is $\delta \omega / \Delta \omega \ll 1$, where $\Delta \omega$ is the difference between the pumped transition frequency and the next closest transition frequency.
This condition is satisfied for the excitations considered in this article.
Here, we review the two-level approximation and present some important results for our analysis.

Consider a nearly monochromatic in-plane electric field with a slowly varying envelope in the graphene mono-layer.
Consider also that the central laser frequency is nearly resonant with the following transition
\begin{equation}
\omega \simeq \omega_{2,-1} = \omega_c (\sqrt{2} + 1).
\end{equation}
According to the RWA, we only keep slowly oscillating (or rotating) terms, i.e. terms with frequencies $\pm|\omega + \omega_{-1,2}|$ in the differential equation for $b_{-1}(t)$ since $| \omega + \omega_{-1,2}| \ll |\omega_{-1,2}|$.
Similarly, we only keep the terms  with frequencies $\pm|\omega - \omega_{2,-1}|$ in the differential equation for $b_{2}(t)$ since $| \omega - \omega_{2,-1}| \ll |\omega_{2,-1}|$.
Neglecting fast oscillating terms, the time-dependent functions entering the expression of the vector potential are, for a RHP excitation (see Appendix \ref{sec:potential})
\begin{equation}\label{eq:f}
G^{\pm}(t) \simeq F^\pm(t) e^{\pm i \omega t},
\end{equation}
where $F^\pm(t)$ is the envelope.
Let us consider the case of zero detuning ($\Delta \equiv \omega - \omega_{2,-1} = 0$), a monochromatic field ($F^\pm = \pm i$) as well as the following initial condition
\begin{equation}\label{eq:init}
\begin{aligned}
b_{-1}(0) &= 1, \\
b_2(0) &= 0,
\end{aligned}
\end{equation}
which corresponds to an initially occupied lower level and initially empty upper level, i.e. a quantum system prepared in a hole-like state.
According to the formulas found in Appendix \ref{sec:analytic2}, the probabilities of the system being in one of the two Landau states at some point in time are given by
\begin{subequations}\label{eq:specific2}
	\begin{align}
	|b_{-1}(t)|^2 &= \cos^2 \left( \frac{\Rabi t}{2} \right), \\
	|b_{2}(t)|^2 &= \sin^2 \left( \frac{\Rabi t}{2} \right).
	\end{align}
\end{subequations}
These oscillations can be interpreted as a periodic change between stimulated absorption and emission of photons by the two-level system \cite{Dora2009}.
In the specific case of zero detuning ($\Delta = 0$), it should be possible to completely fill upper level at the expense of the lower level.
This periodic exchange of energy means that level populations do not generally reach a steady-state value, unless the system is initially in one of its so-called ``dressed states'', which are eigenstates of the dynamical system in the presence of a monochromatic field \cite{Boyd2003}.

\begin{figure}
	\centering
	\includegraphics[]{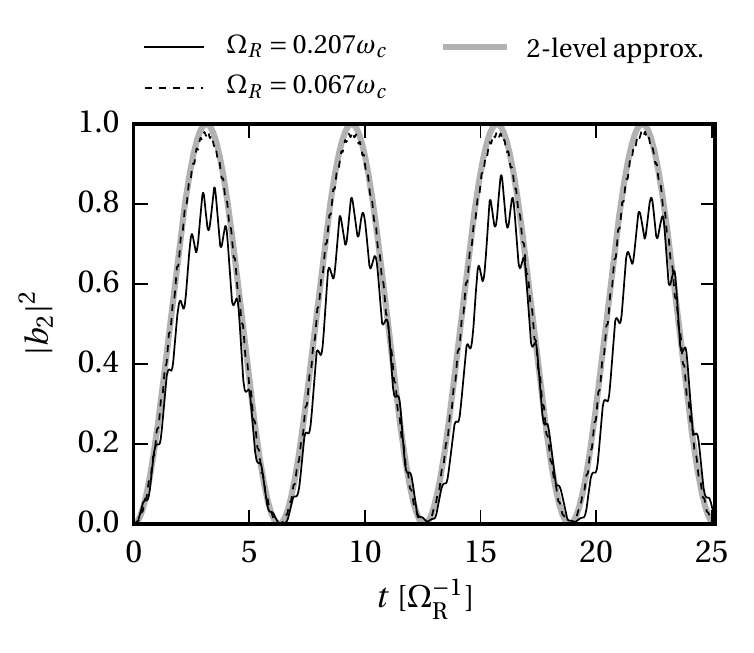}
	\caption{Time evolution of the population of LL +2 for $\Omega_R = 0.207 \omega_c$, corresponding to $E = v_F B$.
		For comparison, a lower Rabi frequency ($\Omega_R = 0.067 \omega_c$) is also shown.
		The monochromatic laser field is tuned to the transition between levels $-1 \rightarrow 2$.
		The closed-form solution obtained from a two-level approximation with zero detuning ($\Delta = 0$) is also shown.
		\label{fig:ccplot}}
\end{figure}

Figure \ref{fig:ccplot} shows the comparison between a numerical result obtained via the full CCM ($\mathcal{N}=30$) and the two-level solution given by Eq. \eqref{eq:specific2}.
The vector potential of the applied monochromatic excitation is defined by Eqs. \eqref{eq:potential} and \eqref{eq:mono}.
This results in a rotating electric field:
\begin{equation}\label{eq:monofield}
\begin{aligned}
E_x & = E_0 \cos (\omega t + \phi), \\
E_y & = E_0 \sin (\omega t + \phi) .
\end{aligned}
\end{equation}
For the sake of demonstration, we set $\phi = 0$ (details in Appendix \ref{sec:potential}).
In these calculations, we have considered the largest magnitude of the in-plane field which does not result in LL collapse in the $\omega \rightarrow 0$ (DC field) limit \cite{Lukose2007}, that is $E = v_F B$ $(\Rabi  = \omega_c^2 / 2 \omega = 0.207 \omega_c$).
In that case, the numerical calculation via the CCM exhibits high frequency oscillations and the peak value of $|b_2|^2$ only reaches about 0.8, which can be attributed to a breakdown of the RWA because of the relatively high Rabi frequency.
Although it is not a ``hard'' limit in the case of an oscillating field, we restrict the discussion to the case $E \lesssim v_F B$ to guarantee that the LL structure is preserved.
The lower the Rabi frequency (i.e. the lower the value of $E$), the lower the difference between the two approaches (two-level vs. CCM), as seen from Fig. \ref{fig:ccplot} for $\Rabi = 0.067 \omega_c$.
For both CCM results, the characteristic frequency of the population oscillation is very close to $\Rabi /2$, as predicted by the two-level solution \eqref{eq:specific2}.
These numerical results obtained via the CCM suggest that the approximate two-level solution should be able, for moderate $\Rabi$, to predict the time scale of Rabi oscillations when considering all possible transition between graphene LLs.
This in turn influences the characteristic time scale of the pair creation process.

\subsection{Validity regime}
\label{sec:assumptions}
We now conclude this section with a discussion of the validity regime of the quantum optical results used in this article.
The main assumption in this work is the absence of many-body effects, i.e. electron-electron interactions.
In single-layer graphene, the strength of the Coulomb interaction is controlled by the coupling parameter $g = e^2 / \kappa \hbar v_F$, where $\kappa$ is the effective dielectric constant of the surrounding medium \cite{Hofmann2014, Basov2014}.
Throughout this paper, $g \ll 1$ is supposed to hold in all calculations.
This condition is clearly not satisfied in suspended graphene ($g \simeq 2.3$).
Consequently, the experimental realization of a weakly coupled device involves embedding the graphene layer in a medium with a sufficiently high dielectric constant $\kappa$.
For example, graphene deposited on SiO$_2$ yields a value of $g \simeq 0.9$ \cite{Basov2014}, and materials with higher dielectric constants are available.

If electron-electron interaction cannot be neglected, the lifetime of hot carriers in graphene is predominantly limited by Auger scattering.
For magnetic fields under 3 T, Mittendorff \emph{et al.} reported that the LL population dynamics exhibited an exponential decay with a time constant $\sim 20$ ps \cite{Mittendorff2014}.
The treatment presented in the previous section can be considered valid if a Rabi period $T_\mathrm{R} \equiv 2 \pi / \Rabi$ is much smaller than 20 ps.
If $B=1$ T and $\Rabi = 0.207 \omega_c$, one obtains $T_\mathrm{R} \simeq 0.5$ ps.
In short, if the condition $g \ll 1$ is not satisfied, one should consider applied electric fields strong enough to ensure a relatively fast Rabi oscillation, but weak enough to preserve the LL structure.

It is worth mentioning a recent article by Wendler \emph{et al.} where electron-electron interactions are not seen as detrimental, but rather exploited to achieve \emph{carrier multiplication} in Landau quantized graphene \cite{Wendler2014}.
This effect could in principle be exploited to increase the number of produced pairs by a factor $\sim 1.3$, but the theoretical treatment of many-body effects in the time-dependent Dirac equation is beyond the scope of the present work.

Besides the aforementioned many-body effects, mechanisms which may reduce the lifetime of excited LLs include electron-phonon scattering and electron-impurity scattering.
Coupling with optical phonons can be mitigated by avoiding tuning the driving laser frequency to optical phonon frequencies, which occupy a relatively narrow energy band \cite{Wendler2014}.
Acoustic phonons, on the other hand, are not expected to significantly impact LL dynamics at very low temperatures \cite{Funk2015}.
Finally, the preferred approach to minimize electron-impurity scattering is to use graphene samples that are as clean as possible \cite{Funk2015}.
Currently available technology allows the production of high quality single crystal layers with domain sizes between 1 and 20 $\mu$m \cite{Basov2014}.

\section{Pair production}
\label{sec:pair}
Before embarking on the mathematical description of pair production, we need to introduce a key feature of LLs: their macroscopic degeneracy.
As detailed in the previous section, the presence of the magnetic field breaks the translational symmetry of the Dirac Hamiltonian, but the physical field is still homogeneous in space.
Classically, this means that the cyclotron orbits corresponding to Landau states can be centered on any given point in space.
Thus, the macroscopic degeneracy of LLs is equal to the number of magnetic flux quanta threading the sample surface, and can be computed by decomposing the position of a Dirac fermion into that of its guiding center (a conserved quantity) and that of a cyclotron trajectory \cite{Goerbig2011}.
In short, the macroscopic degeneracy per surface area of every LL is equal to $n_B = e B/2 \pi$ and thus increases linearly with the magnetic field strength \cite{Gusynin2007}.
Physically, this degeneracy occurs because at higher values of $B$, the cyclotron orbits have a smaller radius $l_B$ and therefore more of them can be packed on the graphene sample.
With this macroscopic degeneracy in mind, one has to introduce a degenerate quantum number, an integer $m$, to completely characterize the system.
The full spinor representation is therefore a tensor product of Hilbert spaces \cite{Goerbig2011}:
\begin{equation}\label{eq:phim}
	\begin{aligned}
	&\ket{\varphi_{0,m; \xi}} = \ket{\varphi_{0; \xi} } \otimes \ket{m}, \\
	&\ket{\varphi_{\lambda n,m; \xi}} = \ket{\varphi_{\lambda n; \xi} } \otimes \ket{m}.
	\end{aligned}	
\end{equation}
If one LL is initially completely empty and subsequently becomes completely filled (which should be possible if an excitation is tuned to an allowed transition), an excess of $4n_B$ charge carriers will be measured in the graphene sample (the factor of 4 comes from the 2 spin branches and the 2 valleys).
This gives an order of magnitude of the number of pairs that could be created by applying a field on a mono-layer.
In this article, we want to answer the following question: can we take advantage of the large macroscopic degeneracy of LLs to bring the produced pair density to a detectable level?
As described by Fillion-Gourdeau and MacLean \cite{Fillion-Gourdeau2015}, the definite answer to that question should come from a many-body approach using second quantized theory.
This theoretical treatment is presented in the next subsection.

\subsection{Second quantization}

Let us introduce the following notation for the discrete states associated with LLs, which form a complete basis.
Define
\begin{equation}
	\ket{\psi_{\lambda n,m; \xi}}= e^{-i \epsilon_{\lambda n} t} \ket{\varphi_{\lambda n,m; \xi}}.
\end{equation}
These discrete states now include the macroscopic degeneracy and satisfy the time-dependent, fully interacting Dirac equation \eqref{eq:dirac2}.
For the remainder of the discussion, we temporarily drop the pseudospin index $\xi$ to facilitate reading.

To obtain an expression for the number of produced pairs, we need to introduce the ``in/out'' formalism from strong field QED.
This allows for an unambiguous definition of asymptotic states \cite[p.~257]{Greiner1985}.
We assume that all observations on the quantum system are made at times long before or long after dynamical changes take place, called ``in/out'' regions.
Suppose that the state of the quantum system is prepared as $\ket{\varphi^{\In}_{\lambda n, m}}$ at $t \rightarrow -\infty$.
Integrating the Dirac equation forward in time (with a possibly time-dependent potential) leads to the following solution, which satisfies the boundary condition \cite[p.~203]{Greiner1985}
\begin{equation}\label{eq:psi+}
	\ket{ \psi^{(+)}_{\lambda n,m} } \xrightarrow{t \rightarrow -\infty} \ket{ \varphi^{\In}_{\lambda n,m} } e^{-i \epsilon^{\In}_{\lambda n} t}.
\end{equation}
Similarly, one can trace the potential changes backwards, which gives
\begin{equation}\label{eq:psi-}
	\ket{ \psi^{(-)}_{\lambda n,m} } \xrightarrow{t \rightarrow \infty} \ket{ \varphi^{\Out}_{\lambda n,m} } e^{-i \epsilon^{\Out}_{\lambda n} t}.
\end{equation}
The (in/out) superscripts refer to the fact that the asymptotic energy levels may be different before and after the field is applied.
This may occur when the dynamical electromagnetic potential is non-zero in asymptotic regions (an example of this is shown in Section \ref{sec:outgoing}).
Let $S_{\lambda n m, \lambda' n' m'}$ be the probability amplitude of a quasi-particle starting in a state $\ket{ \varphi^{\In}_{\lambda' n', m'} }$ and ending in a state $\ket{ \varphi^{\Out}_{\lambda n, m} }$.
According to Greiner \cite{Greiner1985}, this probability amplitude is given by the overlap between solutions of the Dirac equation $\ket{\psi^{(+)}_{\lambda n,m}}$ and $\ket{\psi^{(-)}_{\lambda' n',m'}}$, that is
\begin{equation}
S_{\lambda n m, \lambda' n' m'} = \braket{\psi^{(-)}_{\lambda n,m} | \psi^{(+)}_{\lambda' n',m'}}.
\end{equation}
These scattering matrix elements can be straightforwardly related to the average number of pairs produced from the ``QED vacuum'', a physical observable. 
The number of measured particles in a given electron-like Landau state labeled with quantum numbers $\lambda n m$ ($\lambda n > F$ where $F$ indicates the Fermi level) corresponds to the following expectation value \cite[p.~259]{Greiner1985}
\begin{equation}
N_{\lambda n m}	= \sum_{\lambda' n' < F,m} | S_{\lambda n m, \lambda' n' m'} |^2.
\end{equation}
One can assume that the time-evolution operator does not mix eigenstates labeled by different values of the degenerate quantum number $m$.
In other words, as discussed in Section \ref{sec:assumptions}, we assume that electron scattering mechanisms can be neglected.
Using Eq. \eqref{eq:phim}, one obtains
\begin{equation}
\begin{aligned}
N_{\lambda n m} &= \sum_{\lambda' n' < F,m'}
\left| \braket{m | m'} \braket{\psi^{(-)}_{\lambda n} | \psi^{(+)}_{\lambda' n'}} \right|^2 \\
& = \sum_{\lambda' n' < F}
\left| \braket{\psi^{(-)}_{\lambda n} | \psi^{(+)}_{\lambda' n'}} \right|^2.
\end{aligned}
\end{equation}
Given this result, the total pair production rate can be obtained by summing over all possible final states located above the Fermi level.
We can also re-introduce the valley pseudospin at this point of the derivation and sum over both values of $\xi$. 
For convenience, we use the following normalized pair density in the remainder of the article:
\begin{equation}\label{eq:pprate}
\begin{aligned}
\bar{N} &= \frac{N}{n_B} = \frac{1}{n_B} \sum_{\lambda n > F,m; \xi} N_{\lambda n m} \\
&=  \sum_{\lambda n > F; \xi} \sum_{\lambda' n' < F; \xi}
\left| \braket{\psi^{(-)}_{\lambda n; \xi} | \psi^{(+)}_{\lambda' n'; \xi}}
\right|^2 ,
\end{aligned}
\end{equation}
where $n_B$ is the total surface degeneracy of LLs, accounting for the sum over $m$, and $N$ is the number of pairs per surface area (the measured number is actually $N_\mathrm{tot}\equiv 2N$, taking into account both real spin branches).
Eq. \eqref{eq:pprate} gives the leading order contribution to the pair density, assuming a weak interaction between electrons and holes.
It is similar to previously obtained pair production formulas \cite{Krekora2004, Gelis2006, Fillion-Gourdeau2015, Gelis2015} in that it allows one to compute the average number of pairs created from vacuum by ``preparing'' negative energy states $\ket{ \varphi^{\In}_{\lambda' n'}}$ at $t \rightarrow -\infty$ without an applied field, except for the quantizing $B$-field). These states are then evolved in the presence of a time-dependent electric field and subsequently projected on the outgoing energy states $\ket{\varphi^{\Out}_{\lambda n}}$.
For the purpose of this article, the time-evolution is computed using the CCM and a finite number of ``in'' and ``out'' states are considered, as further described in Section \ref{sec:numerical}.

A final remark remains to be done on the treatment of the ZLL.
We shall treat the ZLL as hole-like for one Dirac point, and electron-like for the other \cite{Gusynin2007}.
In other words, when evaluating the sum given by Eq. \eqref{eq:pprate}, we shall suppose that for one Dirac point, the Fermi level is located between LLs 0 and 1, and that it is located between LLs -1 and 0 for the other Dirac point.
Formally, this can be achieved by the introduction of a Dirac mass term of small magnitude $M$ which shifts the ZLL energy to $\epsilon = -M$ in one valley and to $\epsilon = +M$ in the other.
One can also argue that the Zeeman splitting of LLs (which is much smaller than the level spacing and thus otherwise neglected) justifies the equal sharing of the ZLL by electron and holes \cite{Gusynin2007}.

\subsubsection{Outgoing energy states}\label{sec:outgoing}
Eq. \eqref{eq:potentialt} shows that the value of the vector potential for $t \rightarrow \infty$ is not necessarily zero, which means the basis of outgoing states is not, in general, the same as the basis of incoming states.
The dynamics of the fermions in the outgoing region are governed by the following low energy Hamiltonian
\begin{equation}\label{eq:hamiltonianc}
\ham_{\xi}^\mathbf{B} = \xi \omega_c 
\begin{pmatrix} 0 & \aop - \alpha \\ \aopd + \alpha^* & 0 \end{pmatrix},
\end{equation}
where 
\begin{equation}
\alpha = - \frac{e l_B}{\sqrt{2}} \left(A^\mathbf{E}_x(T) - i A^\mathbf{E}_y(T) \right).
\end{equation}
Since the creation/annihilation operators satisfy the commutation relation $[\hat{a}, \hat{a}^\dagger] = 1$, one can interpret the presence of a non-zero asymptotic potential as a displacement of the free spinors in the $(\Pi_x, \Pi_y)$ phase space.
If one introduces the usual displacement operator \cite{Cahill1969}
\begin{equation}
D(\alpha) = \exp (\alpha \hat{a}^\dagger - \alpha^* \hat{a}),
\end{equation}
then it can be straightforwardly shown (see Appendix \ref{sec:displacement}) that the free spinors of Eq. \eqref{eq:hamiltonianc} are
\begin{equation}\label{eq:zll_shifted}
\ket{\varphi_{0; \xi} } = 
D(\alpha) \begin{pmatrix}
0 \\
\xi \ket{0}
\end{pmatrix},
\end{equation}
for $n = 0$ and
\begin{equation}\label{eq:twospinor_shifted}
\ket{\varphi_{\lambda n; \xi}} = \frac{D(\alpha)}{\sqrt{2}}
\begin{pmatrix}
\ket{n-1} \\
\lambda \xi \ket{n}
\end{pmatrix},
\end{equation}
for $n \neq 0$.
Furthermore, the eigenvalue spectrum of Hamiltonians \eqref{eq:hamiltonianb2} and \eqref{eq:hamiltonianc} is exactly the same, i.e. the displacement operator does not change the energy of the LLs.
Consequently, the Fermi level is the same ($n = 0$) for both incoming and outgoing energy states.

Let us now go back to the calculation of the average number of produced pairs, $\bar{N}$.
Solutions can be related by the time evolution operator $\hat{U}$, which allows one to rewrite
\begin{equation}\label{eq:sandwich}
	\begin{aligned}
		& \braket{\psi^{(-)}_{\lambda n; \xi} | \psi^{(+)}_{\lambda' n'; \xi}} = \\
		& \lim_{\substack{t \rightarrow \infty \\ t' \rightarrow -\infty }} \braket{\varphi^{\Out}_{\lambda n; \xi} | \hat{U}(t,t')| \varphi^{\In}_{\lambda' n'; \xi}}
		e^{i \epsilon^{\Out}_{\lambda n} t -  i\epsilon^{\In}_{\lambda' n'} t'} .
	\end{aligned}
\end{equation}
After the application of the field, the system initially in a pure quantum state is now in a superposition of states given by Eq. \eqref{eq:ansatz}, i.e.
\begin{equation}\label{eq:timeevolution}
\begin{aligned}
&\hat{U}(t,t') \ket{ \varphi^{\In}_{\lambda' n'; \xi}} e^{i\epsilon^{\In}_{\lambda' n'} t'} \\
&= \sum_{\substack{\lambda'' n'' \\ (n'' \neq 0)}} b_{\lambda'' n''; \xi}(t) e^{-i \epsilon_{\lambda'' n''} t } \ket{\varphi_{\lambda'' n''; \xi}} + b_{0; \xi}(t) \ket{\varphi_{0; \xi}}.
\end{aligned}
\end{equation}
Combining equations (\ref{eq:zll_shifted} -- \ref{eq:timeevolution}), setting $t' = 0$, $t= T$ and using the fact that the incoming and outgoing energies are  identical yields
\begin{widetext}
\begin{equation}
\begin{aligned}
\braket{\psi^{(-)}_{\lambda n; \xi} | \psi^{(+)}_{\lambda' n'; \xi}} 
&= \sum_{\substack{\lambda'' n'' \\ (n'' \neq 0)}} b_{\lambda'' n''; \xi}(T) e^{i \omega_c T (\lambda \sqrt{n} - \lambda''\sqrt{n'})} 
\left[ \frac{\bra{n - 1} D (- \alpha)  \ket{n'' - 1}}{2} + \lambda \lambda'' \frac{\bra{n} D (- \alpha)  \ket{n''}}{2}  \right] \\
& + b_{0; \xi}(T)  e^{i \omega_c T \lambda \sqrt{n}} 
\left[ \lambda \frac{\bra{n} D (- \alpha)  \ket{0}}{\sqrt{2}}  \right],
\end{aligned}
\end{equation}
for $n \neq 0$ and
\begin{equation}
\braket{\psi^{(-)}_{\lambda n; \xi} | \psi^{(+)}_{\lambda' n'; \xi}} 
=   \sum_{\substack{\lambda'' n'' \\ (n'' \neq 0)}} \lambda'' b_{\lambda'' n''; \xi}(T) e^{-i \omega_c T  \lambda''\sqrt{n''}} \frac{\bra{0} D (- \alpha)  \ket{n''}}{\sqrt{2}} 
 + b_{0; \xi}(T)\bra{0} D (- \alpha)  \ket{0},
\end{equation}
\end{widetext}
for $n = 0$.
Substituting these results in Eq. \eqref{eq:pprate} allows one to compute the average number of produced pairs for any vector potential using the CCM.
The matrix elements of the displacement operator are proportional to $\exp(-|\alpha|^2)$. 
An explicit form is given in Appendix B of Cahill and Glauber \cite{Cahill1969}.

\section{Numerical results and discussion}

\label{sec:numerical}

\subsection{Monochromatic field}\label{sec:mono}

We now consider the case of an in-plane monochromatic field defined by Eq. \eqref{eq:monofield} (the vector potential is given by Eq. \eqref{eq:mono} in Appendix).
The main objective is to study the impact of the field strength, i.e. the Rabi frequency, on the pair production rate.
The monochromatic excitation is tuned to the transition between LLs -1 and 2, i.e. $\omega = (\sqrt{2} + 1) \omega_c$.
We took $\mathcal{N} = 30$ in the CCM computations and considered 20 initial hole-like states projected on 20 electron-like states to evaluate the sum given by Eq. \eqref{eq:pprate}, which gave a value of $\bar{N}$ accurate to $10^{-3}$ pairs.

\begin{figure}
	\centering
	\includegraphics[]{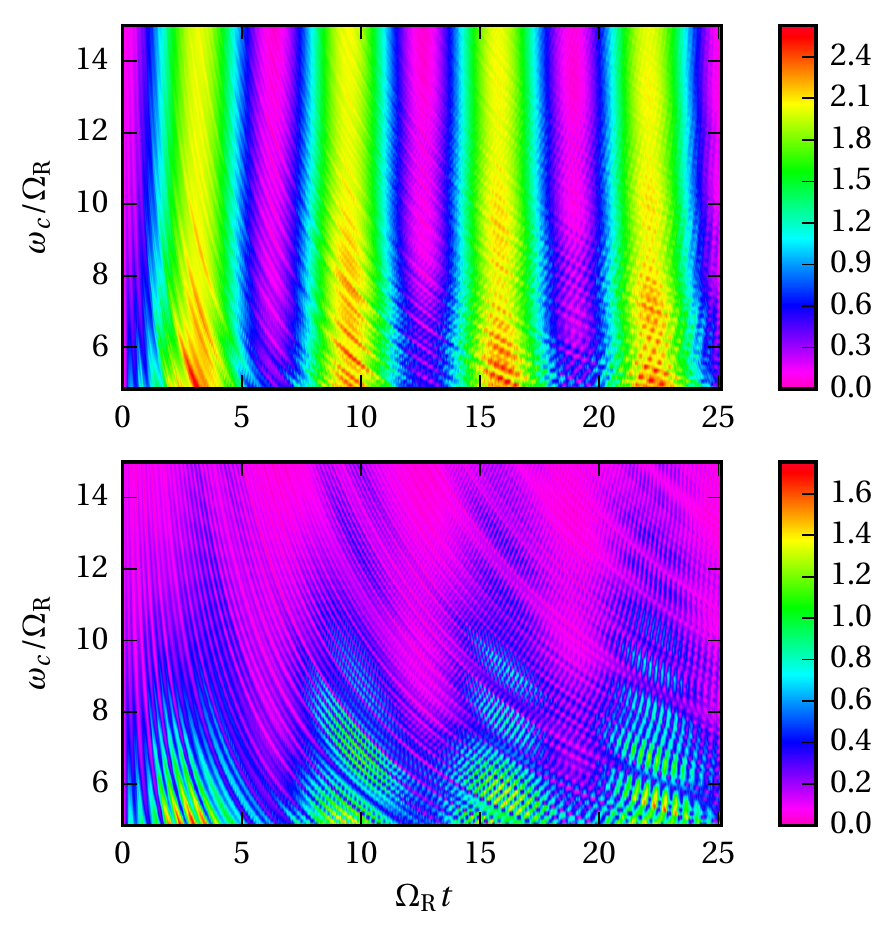}
	\caption{(Top) Dependence of the normalized pair yield $\bar{N}$ on the Rabi frequency for a monochromatic driving field. The monochromatic laser field is tuned to the transition between levels $-1 \rightarrow 2$. The lower bound of the vertical axis corresponds to $E = v_F B$ $(\Rabi = \omega_c^2 / 2 \omega$). (Bottom) Same result excluding pairs produced in the resonant upper LL with energy $\epsilon/\omega_c = \sqrt{2}$.}\label{fig:omegac}
\end{figure}

Figure \ref{fig:omegac} shows the number of produced pairs given by Eq. \eqref{eq:pprate} as a function of the dimensionless parameter $\omega_c / \Omega_R$, i.e. the transition frequency of the first LL divided by the Rabi frequency.
One can see that for large values of $\omega_c / \Omega_R$, i.e. a small magnitude of $E$, the number of produced pairs $\bar{N}$ approaches 2 (one per Dirac point).
This means transitions other than between LLs -1 and 2 contribute negligibly to the sum \eqref{eq:pprate}, and explains why the pair production rate precisely follows a Rabi flopping pattern.
For small values of $\omega_c / \Omega_R$, i.e. a larger applied electric field, $\bar{N}$ can reach values up to 2.5, due to the RWA breaking down and more transitions becoming efficiently driven in an indirect process, i.e. via intermediate dipole-allowed transitions.
This can be confirmed by examining the contribution of the nonresonant upper LLs to the pair density, as shown in Fig. \ref{fig:omegac} (bottom panel).
For $E \simeq v_F B$, the number of pairs produced in other LLs than the resonant level +2 can reach values up to 1.6 and is associated with fast population oscillations, whereas the number of pairs produced in the resonant levels oscillates at the Rabi frequency.
This indirect process cannot be accounted for using a two-level approach.

A physical explanation of this growth in the number of produced pairs is the fact that, for large values of the applied in-plane electric field, the LL spacing is effectively decreased, as described by Lukose \emph{et al.} \cite{Lukose2007}
Thus, the transition rate between non-resonant levels increase. 
If $E = v_F B$ $(\Rabi = \omega_c^2 / 2 \omega$), the LLs experience a collapse phenomenon in the $\omega \rightarrow 0$ (DC field) limit: they merge with each other and form a continuum.

The oscillatory behavior shown in Fig. \ref{fig:omegac} suggests that, for a given magnitude of the applied electric field $E$, it should be possible to maximize the pair yield by choosing an appropriate pulse envelope, a procedure known in the quantum computing community as applying a ``$\pi$-pulse'' to completely invert a two-level atom, or qubit \cite{Biolatti2000}.
This hypothesis is verified numerically in the next subsection by taking all possible graphene LL transitions into account.

\subsection{Slowly varying envelope}
 
Consider now the following RHP excitation, applied between $0 < t < T$, with $T = \pi / a$:
\begin{equation}
\begin{aligned}
E_x & = E_0 \sin^2 at \cos \omega t , \\
E_y & = E_0 \sin^2 at \sin \omega t .
\end{aligned}
\end{equation}
The main objective is to evaluate the influence of the pulse duration in terms of carrier cycles, i.e. $T/ T_0 = \omega / 2 a$, on the pair yield.
In other words we seek the values of $T$ that maximize the number of produced pairs.
Once again, the carrier frequency is tuned to the transition between LLs -1 and 2.
CCM results are computed for various values of $T / T_0$.
We take $\mathcal{N} = 50$ in the CCM computations and consider 24 initial hole-like states projected on 24 electron-like states to evaluate the sum given by Eq. \eqref{eq:pprate}, which ensures a value of $\bar{N}$ numerically accurate to $10^{-3}$ pairs.

\begin{figure}
	\centering
	\includegraphics[]{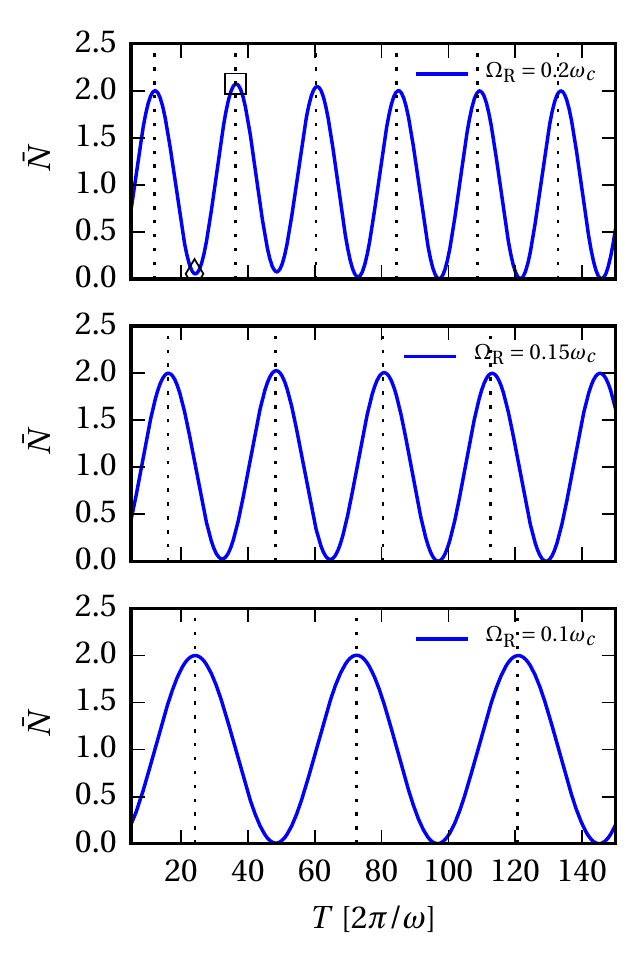}
	\caption{Dependence of the pair yield on the driving pulse duration. The location of the maxima predicted by the two-level approximation, i.e. Eqs. \eqref{eq:analytic} and \eqref{eq:ftilde}, are indicated by dashed lines (odd integer values of $\Rabi T / 2 \pi$).}\label{fig:widthplot}
\end{figure}

Figure \ref{fig:widthplot} shows the numerically evaluated pair yield as a function of the pulse duration, for 3 different values of the Rabi frequency.
One can see that local maxima in the number of produced pairs follow a periodic pattern, with maxima corresponding almost exactly to odd integer values of $\Rabi T / 2 \pi$.
As the Rabi frequency decreases, the contribution of the non-driven levels to the pair density decreases as well.
These results show that in the presence of a slowly varying envelope, the transition pumped by the carrier wave still dominates the dynamics, even though other levels may participate in pair production, as exemplified by values of $\bar{N}$ slightly higher than 2 (especially for high Rabi frequencies).
In short, the carrier frequency provides control over the transition one wants to select, whereas the pulse duration provides control over the final population of the upper level.

\begin{figure}
	\centering
	\includegraphics[]{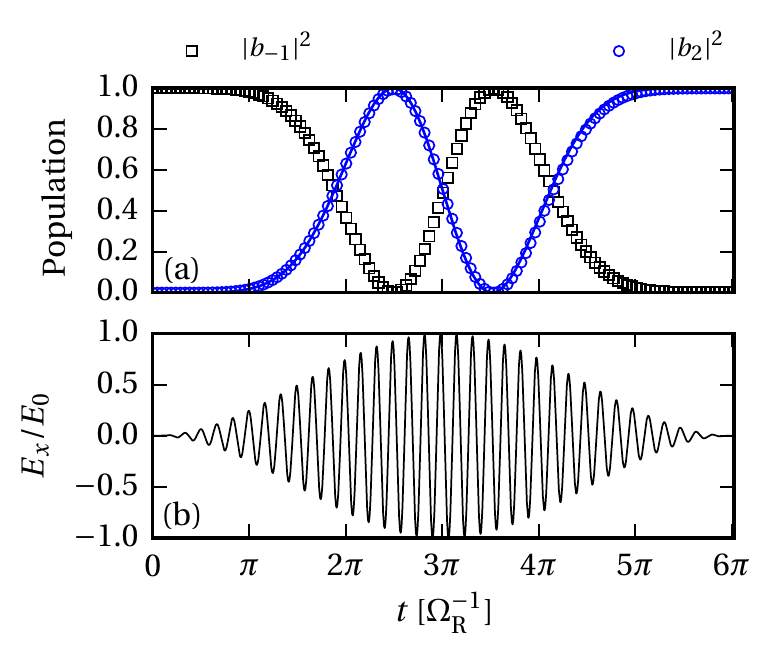}
	\caption{(a) LL dynamics with a pulsed field with $T/T_0 = 36.33$, corresponding to the square marker on Fig. \ref{fig:widthplot}.
		This corresponds to a value which maximizes the population of LL 2 after the passage of the pulse, thus maximizing the pair yield.
		The markers indicate the multi-level numerical result obtained via CCM, and the blue curve indicates the approximate two-level solution.
		(b) Driving pulse profile.}\label{fig:yield}
\end{figure}

\begin{figure}
	\centering
	\includegraphics[]{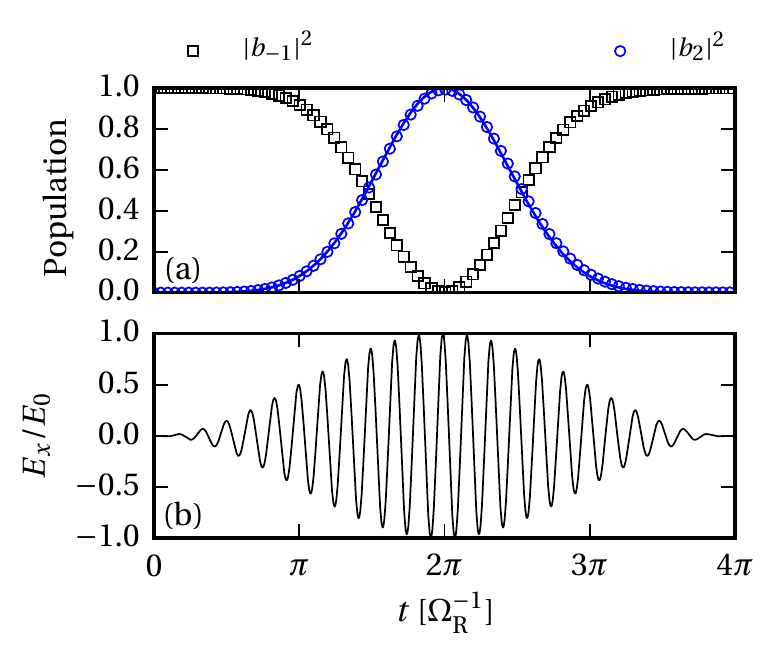}
	\caption{(a) LL dynamics with a pulsed field with $T/T_0 = 24.0$, corresponding to the diamond marker on Fig. \ref{fig:widthplot}.
	This corresponds to a local minimum of the population of LL 2 after the passage of the pulse.
	The markers indicate the multi-level numerical result obtained via CCM, and the blue curve indicates the approximate two-level solution.
	(b) Driving pulse profile.}\label{fig:yield_low}
\end{figure}

Let us further the analysis of the multi-level CCM results of Fig. \ref{fig:widthplot}, concentrating on the results for $\Omega_\mathrm{R} = 0.2 \omega_c$.
Fig. \ref{fig:yield} shows the time-evolution of the level populations participating in the driven transition (i.e. LLs -1 and 2) for $T/ T_0 = 36.33$, which corresponds to the value giving the highest pair yield ($\bar{N} \simeq 2.073$).
One can see that the population of the upper level is very close to 1 after the passage of the pulse.
The additional $0.073$ pairs (see square marker on Fig. \ref{fig:widthplot}) are the contribution of the other transitions, owing to the relatively high value of the Rabi frequency.
For comparison purposes, the case of a pulse width which minimizes the population of the upper level is shown in Fig. \ref{fig:yield_low} (corresponding to the diamond marker in Fig. \ref{fig:widthplot}).
In this case, $\bar{N} \simeq 0.05$.

For completeness, numerical results obtained via CCM are compared to curves (shown in Fig. \ref{fig:yield} and \ref{fig:yield_low}) obtained from the approximate two-level solution.
The relatively good agreement between the two approaches shows that the two-level approximation may be used to predict the pulse duration needed to completely invert the level population, even for large Rabi frequencies.

To conclude this discussion, we give representative numerical values of the pair density for the situation that maximizes $\bar{N}$ (see Table \ref{tab:parameters}).
Since $n_B = e B / 2 \pi$, the pair density $N_\mathrm{tot}$ is fixed by the value of the magnetic field.
For $B = 0.01$ T, we have $N_\mathrm{tot} \sim 10^{13} \mbox{ m}^{-2}$, while $B = 100$ T gives $N_\mathrm{tot} \sim 10^{17}$ $\mbox{m}^{-2}$.
The former value corresponds to a magnetic field that can be produced using a static magnet, while the latter could be achieved via a strain induced gauge field \cite{Levy2010, Zhu2015}.
Since the typical LL spacing increases with the magnetic field, the frequency and intensity of the driving field have to be adjusted accordingly.
A magnetic field of $B = 0.01$ T implies using a $\sim$ 2 THz laser source, while a magnetic or pseudo-magnetic field of $B = 100$ T implies a laser operating near 200 THz.
The pair density for $B=100$ T interestingly reaches up to $10^{17}$ m$^{-2}$, which is orders of magnitude higher than the values obtained in the non-magnetized case \cite{Fillion-Gourdeau2015}.
However, the experimental configuration corresponding to the third row of Table \ref{tab:parameters} is challenging to achieve.
Laser intensities lower than $10^{10}$ W/cm$^2$ should also be employed to preserve the integrity of the sample \cite{Roberts2011}; this would only affect the Rabi frequency with minimal impact on the pair yield.

The markedly higher values of $N_\mathrm{tot}$ found in Table \ref{tab:parameters} can be explained by the macroscopic degeneracy of LLs and their resonant behavior, both features exclusive to magnetized graphene.
In short, the macroscopic degeneracy provides an additional quantum number which can be exploited to sidestep the Pauli exclusion principle. 
This blocking limits the achievable pair densities in the non-magnetized case \cite{Su2012, Fillion-Gourdeau2015}.

\begin{table}
	\caption{Representative numerical values of the magnetic field, driving field parameters and resulting pair density corresponding to the
		maximum value of $\bar{N}$ (square marker on Fig. \ref{fig:widthplot}, $\Rabi = 0.2 \omega_c$).
	The last column corresponds to numerical values obtained via the CCM.}\label{tab:parameters}
	\begin{center}
		\begin{tabular}{ccc|c}
			\hline
			$B$ (T) & $\nu_0$ (THz) & $I$ (W/cm$^2$) & $N_\mathrm{tot}$ (m$^{-2}$) \\ \hline
			0.01 & 2.12 & $4.95 \times 10^1 $ & $1.003 \times 10^{13}$ \\
			1 & 21.2 & $4.95 \times 10^5$ & $1.003 \times 10^{15}$ \\
			100 & 212 & $4.95 \times 10^{9}$ & $1.003 \times 10^{17}$ \\
			%300 & 367 & $4.45 \times 10^{10}$ & $3.01 \times 10^{17}$ \\ \hline
		\end{tabular}
	\end{center}
\end{table}

It should be noted that the results obtained in this work are markedly different from usual QED studies which considered crossed \cite{Su2012} or even collinear \cite{Tanji2009} static electric and magnetic fields. 
These studies conclude that the presence of the magnetic field results in a suppression of the pair production rate, rather than an enhancement, because it makes fermions ``heavy''.
The apparent discrepancy with the present graphene QED study is due to the fact that we have only considered relatively weak electric fields with magnitudes $E < v_F B$, whereas other authors considered $E \geq v_F B$ (or $E \geq c B$ for the case of usual QED).
In the case of $E \geq v_F B$ and $\omega \rightarrow 0$, LLs do not exist and it is possible to apply a Lorentz boost to the Dirac equation which eliminates the magnetic field and effectively reduces the value of the electric field \cite{Lukose2014}.
Since this situation results in a suppression of the pair production rate, we did not consider it in the present work.
We also did not consider the effect of strong magnetic fields parallel to the graphene layer, since it does not result in Landau quantization.

\subsection{Linear polarization}\label{sec:linear}

We now briefly discuss the impact of using linearly instead of circularly polarized laser excitations to drive pair creation with graphene LLs.
Consider first a linearly polarized monochromatic field
\begin{equation}\label{eq:linear}
E_x = 2 E_0 \cos \omega t, \qquad E_y = 0.
\end{equation}
This excitation can be decomposed as a superposition of a RHP and a LHP excitation (see Eq. \ref{eq:monofield}).
The twice higher field ($2 E_0$) is chosen so that the amplitude of the two orthogonal components is $E_0$.
As a result, each excitation drives the LLs with the same Rabi frequency as the single-component circularly polarized case presented in section \ref{sec:mono}.

\begin{figure}
	\centering
	\includegraphics[]{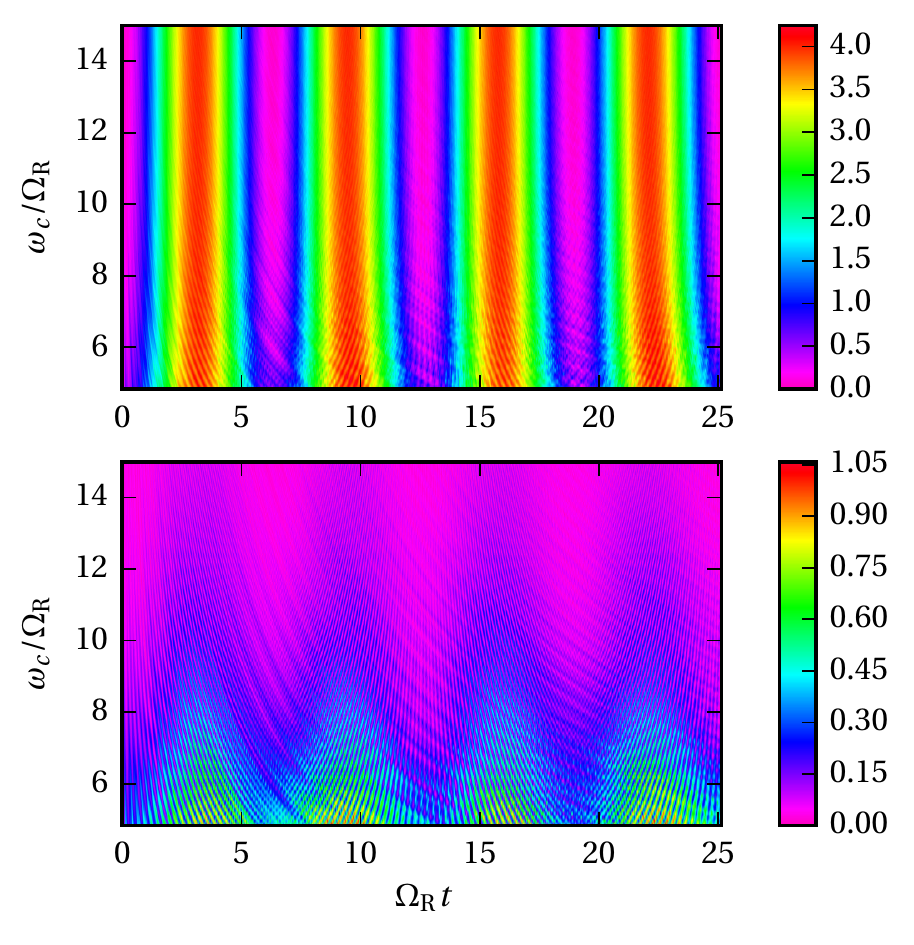}
	\caption{(Top) Dependence of the normalized pair yield $\bar{N}$ on the Rabi frequency for a linearly polarized monochromatic driving field. The monochromatic laser field drives the transition between levels $-1 \rightarrow 2$ and the transition $-2 \rightarrow 1$. The lower bound of the vertical axis corresponds to $E = v_F B$ $(\Rabi = \omega_c^2 / 2 \omega$). (Bottom) Same result excluding pairs produced in the resonant upper LLs with energies $\epsilon/\omega_c = 1$ and $\epsilon/\omega_c = \sqrt{2}$.}\label{fig:omegac_linear}
\end{figure}

CCM computations are performed for a linearly polarized excitation with the same parameters found in Section \ref{sec:mono}.
As can be seen from Fig. \ref{fig:omegac_linear}, a Rabi flopping pattern with the same frequency as the circular polarization case is obtained.
The key difference is that the pair yield is increased approximately by a factor of 2.
This result can be explained by treating each of the circular polarization components independently.
Applying the RWA shows that the RHP component efficiently drives the transition between LLs -1 and 2, whereas the LHP component drives the transition between LLs -2 and 1, as indicated in Fig. \ref{fig:transitions}.
Accordingly, the net effect of considering a linearly polarized excitation is that two transitions of equal frequency instead of one can contribute to the pair yield.
This result suggests that linearly polarized lasers could be advantageously used in experiments to bring the graphene pair yield to a detectable level.
It also shows the possibility of driving multiple transitions at the same time, which could be realized by using different laser colors simultaneously.

\begin{figure}
	\centering
	\includegraphics[]{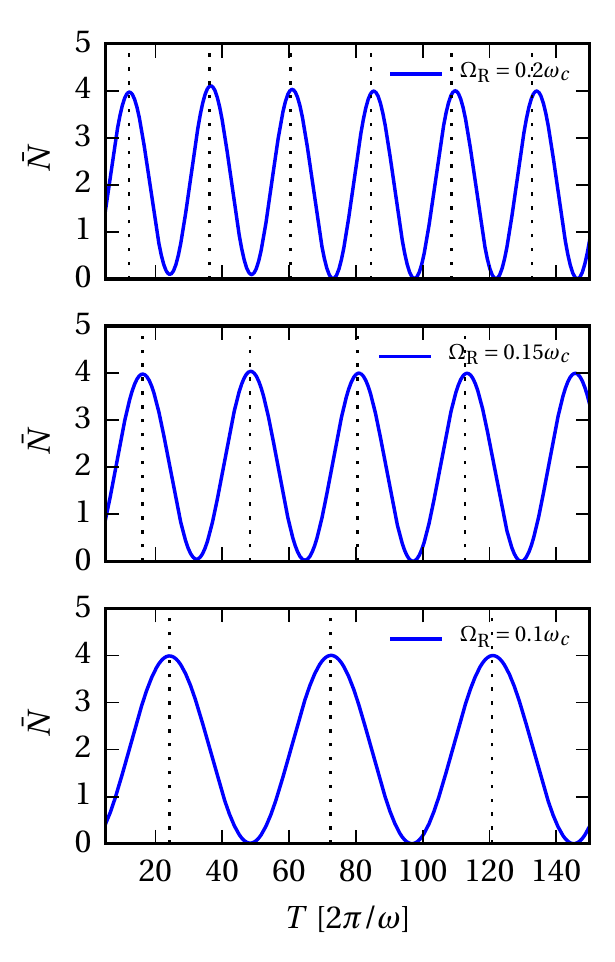}
	\caption{Dependence of the pair yield on the driving pulse duration (linear polarization). The location of the maxima predicted by the two-level approximation, i.e. Eqs. \eqref{eq:analytic} and \eqref{eq:ftilde}, are indicated by dashed lines (odd integer values of $\Rabi T / 2 \pi$).}\label{fig:widthplot_linear}
\end{figure}

Similar results are obtained for the case of the time-varying envelope, as seen in Fig. \ref{fig:widthplot_linear}.
In that case, the field is given by
\begin{equation}\label{eq:linear_sine}
E_x = 2 E_0 \sin^2 a t \cos \omega t, \qquad E_y = 0,
\end{equation}
and is applied for $T = \pi / a$.

\section{Conclusion and outlook}\label{sec:conclusion}

In this work, quantum tunneling between graphene LLs driven by a time-dependent electric field was investigated theoretically and numerically.
We considered weak driving fields ($E < v_F B$) and the case of linear and circular polarization orthogonal to the $B$-field direction.
The coupled channel method (CCM) was used to solve the Dirac equation in the presence of the quantizing $B$-field and the driving field, with solutions interpreted in terms of a produced pair density.
This time-dependent pair creation demonstrates that magnetized graphene can be employed as a QED analogue, and that the high macroscopic degeneracy and resonant behavior of LLs can be exploited to increase the pair yield with respect to the non-magnetized case.
Although we considered excitation frequencies tuned to a single transition between LLs (in which case the dynamics can be described by a two-level approximation in a satisfactory way), the numerical method used in the article is able to deal with a generic time-dependent excitation and any number of coupled levels.

Experiments to probe the dynamics of LLs could realistically be carried using moderate intensity laser sources ranging from 10--400 THz, depending on the applied magnetic field strength.
Pair densities ranging from $10^{14}$ -- $10^{17}$ m$^{-2}$ could theoretically be obtained, although the experimental detection of the produced pairs remains a challenge because of the finite lifetime of hot carriers in graphene.
Linearly polarized lasers could be advantageously used to maximize the pair yield owing to the peculiar selection rules of magnetized graphene, i.e. the fact that every allowed transition frequency is associated to \emph{two} LL transitions of different handedness.
Further work could include pair production using a DC electric field applied to a magnetized layer, a situation which was not considered in this paper.
A DC field has the effect of ``tilting'' the Dirac cones, which is associated to a multiplication of the number of allowed transitions, as described by Sari \emph{et al.} \cite{Sari2015} 

Using a simple sine-squared pulse model, we demonstrated numerically that the driving pulse duration provides a control parameter over the process of pair creation in magnetized graphene.
This result suggests that the optimization of the spectral content of the incident pulse should allow one to maximize the pair yield by driving multiple LL transitions simultaneously.
In fact, pulse shaping was used in a recent theoretical paper to optimize pair production from vacuum using ultra-intense lasers, albeit in the absence of a magnetic field \cite{Hebenstreit2014}.

%\section*{Acknowledgements}
\begin{acknowledgments}
Computations were made on the supercomputer \emph{Mammouth}, managed by Calcul Qu\'ebec and Compute Canada.
The operation of this computer is funded by the Canada Foundation for Innovation (CFI), Minist\`ere de l'\'Economie, de l'Innovation et des Exportations du Qu\'ebec (MEIE), RMGA and the Fonds de recherche du Qu\'ebec - Nature et technologies (FRQNT).
\end{acknowledgments}

\bibliography{cond-mat}

\appendix

\section{Description of the coupled channel method}
\label{sec:ccm}
The coupled channel method (CCM) used for solving Eqs. \eqref{eq:ode3} is summarized in this Appendix.
The ordinary differential equation (ODE) system is exact provided we include an infinite number of levels; this is however not possible \cite{Wang2008}.
First we consider a truncated version of the ODE system \eqref{eq:ode3} with valence band levels $n \in [1, \mathcal{N}]$, conduction band levels $n \in [1, \mathcal{N}]$ and the ZLL, for a total number of levels $2 \mathcal{N} + 1$.
This results in a closed ODE system, which can then be numerically solved using standard integration routines: we use an explicit Runge-Kutta method of order 8 provided by the Python package Scipy \cite{Jones2001}. A relative error tolerance of $10^{-6}$ is employed, resulting in a converged solution.

The value of $\mathcal{N}$ must be chosen large enough to ensure convergence: this happens when energy levels higher than $\mathcal{N}$ (or lower than $-\mathcal{N}$) are not populated or do not participate in the dynamics.
If the levels with index $n \geq \mathcal{N}$ are non-resonant, this should be the case to a good degree.
It is possible to set the value of $\mathcal{N}$ according to the criterion
\begin{equation}
\mathcal{N} \sim \beta^2 \left[ \frac{\Rabi^2}{\omega_c^2} \right],
\end{equation}
which is equivalent to requiring that the energy of the highest LLs considered is of order $\beta \Rabi$, where $\beta$ is some proportionality constant chosen to ensure convergence.

\section{Solution in the two-level approximation} \label{sec:analytic2}
The goal of this Appendix is to obtain an approximate, closed form solution of system \eqref{eq:ode3}, under the RWA, and considering only two driven levels.
Under these conditions, starting from Eqs. \eqref{eq:ode3}, one obtains the following differential equations for a closed two-level system (without loss of generality, we select the transition between LLs -1 and 2 and suppose a RHP excitation):
\begin{subequations}\label{eq:rabi}
	\begin{align}
	&\dot{b}_{-1}(t) = \frac{i \Rabi}{2} b_2(t) F^+(t) e^{i \Delta t}, \\
	&\dot{b}_{2}(t) = \frac{i \Rabi}{2} b_{-1}(t) F^-(t) e^{-i \Delta t},
	\end{align}
\end{subequations}
where the detuning factor is defined as
\begin{equation}
\Delta \equiv \omega + \omega_{-1,2} = \omega - \omega_{2,-1}.
\end{equation}
Consider the change of variables $b_{-1}(t) = c_{-1}(t) e^{i \Delta t}$, $b_{2}(t) = c_{2}(t) e^{-i \Delta t}$.
The system of equations \eqref{eq:rabi} can be recast in matrix form as
\begin{equation}\label{eq:matrix1}
\dot{\mathbf{c}}(t) = 
\begin{bmatrix}
\frac{-i \Delta}{2} & \frac{i \Omega_R}{2} F^+(t) \\
\frac{i \Omega_R}{2} F^-(t) & \frac{i \Delta}{2} 
\end{bmatrix} \mathbf{c}(t)  = M(t) \mathbf{c}(t)
\end{equation}
where $\mathbf{c}(t) = \left(c_{-1}(t),c_{2}(t) \right)^\mathrm{T}$.
The formal solution to this system of equations is given by a time-ordered exponential \cite{Blanes2009}
\begin{equation}\label{eq:matrix2}
\mathbf{c}(t) = \mathcal{T} \exp \left[  \int_0^t M(t') dt' \right] \mathbf{c}(0).
\end{equation} 
For the purposes of this article, we shall approximate the time-ordered exponential by its Magnus expansion \cite{Blanes2009}.
In a nutshell, the Magnus expansion allows one to rewrite the time-ordered exponential as a true matrix exponential, the argument of which is an infinite series.
Truncating this series to its leading term simply amounts to removing the time-ordering symbol:
\begin{equation}\label{eq:matrix3}
\mathbf{c}(t) = \exp \left[  \int_0^t M(t') dt' \right] \mathbf{c}(0).
\end{equation} 
Combining Eqs. \eqref{eq:matrix1} and \eqref{eq:matrix3}, we obtain the following approximate solution for the two-level system
\begin{equation}
\mathbf{c}(t) = \exp
\begin{bmatrix}
\frac{-i \Delta t}{2} & \frac{i \Omega_R}{2} \tilde{F}^+(t) \\
\frac{i \Omega_R}{2} \tilde{F}^-(t) & \frac{i \Delta t}{2} 
\end{bmatrix} \mathbf{c}(0),
\end{equation} 
where $\tilde{F}^\pm(t) = \int_0^t F^\pm(t')dt'$.
The matrix exponential of a $2 \times 2$ matrix can be computed analytically.
For the specific initial condition \eqref{eq:init} and zero detuning $\Delta = 0$, one has \cite{Exponential}
\begin{equation}\label{eq:analytic}
\begin{aligned}
c_{-1}(t) & = \cos\left( \frac{\Omega_R}{2} \sqrt{\tilde{F}^+(t) \tilde{F}^-(t) } \right), \\
c_{2}(t) & = i \sqrt{\frac{\tilde{F}^+(t)}{\tilde{F}^-(t)}}\sin\left( \frac{\Omega_R}{2} \sqrt{\tilde{F}^+(t) \tilde{F}^-(t) } \right),
\end{aligned}
\end{equation}
and $|b_{-1}(t)|^2 = |c_{-1}(t)|^2$, $|b_{2}(t)|^2 = |c_{2}(t)|^2$.
It should be noted that this approximate solution is exact for any function $F^{\pm}(t)$ that is a constant since $[M(t),M(t')] = 0$ under this condition.
This is the case for a monochromatic excitation.

\section{Explicit form of the vector potential}\label{sec:potential}
One needs to obtain an expression for the vector potential
\begin{equation}\label{eq:potential2}
\mathbf{A}^\mathbf{E}(t) = A_0 \left\lbrace G_x(t) \hat{\mathbf{e}}_x + G_y(t)\hat{\mathbf{e}}_y \right\rbrace
\end{equation}
starting from the expression of the electric field
\begin{equation}\label{eq:efield}
\mathbf{E}(t) = -\frac{d \mathbf{A}^\mathbf{E}(t)}{d t} = -A_0 \left\lbrace \frac{d G_x(t)}{dt} \hat{\mathbf{e}}_x + \frac{d G_y(t)}{dt}\hat{\mathbf{e}}_y \right\rbrace.
\end{equation}
Since only $G^\pm(t) = G_x(t) \pm i G_y(t)$ enters in the differential equations, one needs to integrate the following expression starting from Eq. \eqref{eq:efield}
\begin{equation}\label{eq:integrate}
\frac{d G^\pm (t)}{dt} = \frac{d G_x(t)}{dt} \pm i \frac{d G_y(t)}{dt} = -\frac{1}{A_0} \left( E_x \pm i E_y \right).
\end{equation}
\subsection{Monochromatic field}
Consider a RHP monochromatic excitation. 
One has
\begin{equation}
\begin{aligned}
E_x & = E_0 \cos (\omega t + \phi), \\
E_y & = E_0 \sin (\omega t + \phi) .
\end{aligned}
\end{equation}
Using Eq. \eqref{eq:integrate}, De Moivre's identity, and the condition $\mathbf{A}^{\mathbf{E}}(0) = 0$, one finds
\begin{equation}\label{eq:mono}
G^{\pm} = \pm i(e^{\pm i\omega t}e^{\pm i \phi} - 1).
\end{equation}
with $A_0 = E_0 / \omega$. 
According to the RWA, we keep only the oscillating field component, that is $F^\pm \simeq \pm i e^{\pm i \phi}$.
One subsequently has $\tilde{F}^\pm = \pm i te^{\pm i \phi}$, and plugging this result in Eqs. \eqref{eq:analytic} yields \eqref{eq:specific2}.

\subsection{Slowly varying envelope}
Consider the following field, applied for $0 < t < \pi / a$:
\begin{equation}
\begin{aligned}
E_x & = E_0 \sin^2 at \cos (\omega t + \phi), \\
E_y & = E_0 \sin^2 at \sin (\omega t + \phi) .
\end{aligned}
\end{equation}
Using Eq. \eqref{eq:integrate} and $A_0 = E_0 / \omega$, one finds
\begin{equation}
G^\pm (t) = - \omega e^{\pm i \phi} \int  \sin^2 at ~ e^{\pm i \omega t} dt.
\end{equation}
Integrating by parts and choosing the integration constant such that $G^\pm(0) = 0$ yields
\begin{widetext}
\begin{equation}
G^\pm (t) =  
\dfrac{\pm i (4a ^2 - \omega^2) \pm i \omega^2 \cos 2 a t + 2 a \omega \sin 2 a t}{2 (4 a^2 - \omega^2)} e^{\pm i \phi} e^{\pm i \omega t}
\mp \dfrac{2ia^2 e^{\pm i \phi}}{4 a^2 - \omega^2},
\end{equation}
and the value of $\tilde{F}$ is 
\begin{equation}\label{eq:ftilde}
\tilde{F}^\pm (t) =
\dfrac{\pm i t (4a ^2 - \omega^2) \pm i \frac{\omega^2 }{2a} \sin 2 a t +  \omega (1-\cos 2 a t)}{2 (4 a^2 - \omega^2)} e^{\pm i \phi}.
\end{equation}
\end{widetext}
Substituting this result in Eq. \eqref{eq:analytic} shows that the level population is insensitive to the carrier-envelope phase $\phi$ (in the validity limit of the RWA and the truncated Magnus expansion) as is the case for the monochromatic excitation.

\section{Properties of the displacement operator}\label{sec:displacement}
In this appendix, we give some useful properties of the displacement operator
\begin{equation}
D(\alpha) = \exp (\alpha \hat{a}^\dagger - \alpha^* \hat{a}),
\end{equation}
and use them to obtain the spinors given by Eq. \eqref{eq:twospinor_shifted}.
The displacement operator is unitary
\begin{equation}
D^{\dagger}(\alpha) = D(-\alpha)
\end{equation}
and satisfies the following relations \cite{Cahill1969}
\begin{align}
&\hat{a}^\dagger - \alpha^* = D(\alpha) \hat{a}^\dagger D(-\alpha), \\
&\hat{a} - \alpha = D(\alpha) \hat{a} D(-\alpha).
\end{align}
Using these identities, it is possible to write down an eigenvalue equation for the spinor component $\varphi_{B,+}$.
Starting from the Hamiltonian \eqref{eq:hamiltonianc} and concentrating on the $K^+$ valley, we obtain
\begin{equation}
D(\alpha)\hat{a}^\dagger \hat{a} D(-\alpha) \varphi_{B,+} = (\epsilon / \omega_c)^2 \varphi_{B,+}.
\end{equation}
Up to a phase factor:
\begin{equation}
\varphi_{B,+} = \ket{\alpha, n} = D(\alpha) \ket{n}.
\end{equation}
Thus
\begin{equation}
 n D(\alpha) \ket{n} = (\epsilon / \omega_c)^2  D(\alpha) \ket{n},
\end{equation}
and
\begin{equation}
\epsilon_{\lambda, n} = \lambda \sqrt{n} \omega_c.
\end{equation}
A similar treatment yields the value of $\varphi_{A,+}$, and can be repeated around the $K^-$ point to obtain Eq. \eqref{eq:twospinor_shifted}.

\end{document}